\documentclass[12pt,a4paper]{article}

\usepackage{comment}

\date{}

\title{\textbf{
Reduced Order Podolsky Model
}}
\author{ \textbf{
Ronaldo Thibes
}
\\
\textit{\small{Departamento de Ci\^encias Exatas e Naturais}}\\
\textit{\small{Universidade Estadual do Sudoeste da Bahia}}\\
\textit{\small{Rodovia BR 415, Km 03, S/N}}\\
\textit{\small{Itapetinga - Bahia - Brazil}}
 }

\usepackage{graphicx}
\usepackage{amssymb,amsfonts,amssymb,amsthm}
\usepackage{amsmath}

\begin{document}

\maketitle

\abstract{
We perform the canonical and path integral quantizations of a lower-order derivatives model describing Podolsky's generalized electrodynamics.
The physical content of the model shows an auxiliary massive vector field coupled to the usual electromagnetic field.  The equivalence with Podolsky's
original model is studied at classical and quantum levels.  Concerning the dynamical time evolution we obtain a theory with two first-class and
two second-class constraints in phase space.  We calculate explicitly the corresponding Dirac brackets involving both vector fields.
We use the Senjanovic procedure to implement the second-class constraints and the Batalin-Fradkin-Vilkovisky path integral quantization scheme
to deal with the symmetries generated by the first-class constraints.  The physical interpretation of the results turns out to be simpler due to
the reduced derivatives order
permeating the equations of motion, Dirac brackets and effective action.
}

\section{Introduction}
Back in 1942
Boris
Podolsky
\cite{Podolsky:1942zz}
introduced 
a second-order derivatives Lagrangian which came to be known as a
generalized electrodynamics \cite{Podolsky:1944zz, Podolsky:1948}.
The resulting equations of motion are of fourth-order in the derivatives of the gauge field.
With an additional length parameter $a$ the initial motivation was related to the removal of infinities coming from
the treatment of point charges.
In his original paper \cite{Podolsky:1942zz}, Podolsky argued that the only way to generalize Maxwell-Lorentz
electrodynamics maintaining linearity was through the introduction of higher-order derivatives, ``{\it unless one is prepared
to introduce new kinds of field quantities}'' \cite{Podolsky:1942zz}.
It seems indeed not only should we be prepared to do so but that both procedures are, to a certain extension, equivalent.
In this present paper we show that Podolsky's model is equivalent to a reduced-order derivatives one, with an extra auxiliary vector field.
The extra field is responsible for the known massive excitations of the theory and the latter turns out to be more attractive and manageable due to the simplicity
arising from lower-order derivatives.
Therefore both approaches complement each other permitting new ways of interpreting the description of the same physical phenomena.

There exists nowadays an ingoing increasing general interest in higher-derivative models in field theory, either from the applied point of view or 
from the fundamental one.
Due to its generality appeal, the understanding of higher-order theories constitutes a fascinating challenge to physicists and mathematicians.
In particular, recently, Podolsky's generalized electrodynamics has been revisited and scrutinized in its
various aspects in a handful of papers \cite{Bufalo:2012tt, Bufalo:2013joa, Zayats:2013ioa, Bufalo:2014jra, Gratus:2015bea, Barone:2016pyy},
establishing a lively rich discussion on the subject.
The model may be used as an effective theory itself or as a smaller component of more elaborated ones.  Considered as a fundamental theory one
has to deal with the fact that up to now no massive photons have been observed.
However, due to technical limitations and the very nature of measurement, only maximal values can be set to such a possible nonzero photon mass.
A more prominent recent discussion of the experimental aspects
related to massive photons in the context of Podolsky's electrodynamics can
be seen for instance in \cite{Cuzinatto:2011zz}. 

Podolsky's generalized electrodynamics is an Abelian gauge theory, whose action contains second-order derivatives of the gauge field, both in
time and space in a covariant way -- the equations of motion are fourth-order partial differential equations.  The Dirac-Bergmann constraint
analysis was performed almost thirty years ago by Galvao and Pimentel \cite{Galvao:1986yq}.  Due to the intrinsic higher-order derivatives
content of the model, the Hamiltonian time evolution had to be described by means of a generalized Legendre transformation involving an extra set
of canonical momenta conjugated to the second time derivative of the gauge field.
That means essentially that in configuration space one expresses fourth-order time derivatives of the fields as functions of the corresponding lower-order derivatives.
In this sense, $A_\mu$, ${\dot A}_\mu$ and ${\ddot A}_\mu$ are considered as independent variables and when passing
to the phase space there must be a counterpart with extra momenta.
Proceeding this way in \cite{Galvao:1986yq}, Galvao and Pimentel found a rich Hamiltonian structure involving three first-class
constraints and calculated the resulting Dirac brackets after gauge fixing.  The functional quantization of the same model was discussed three years
later by Barcelos-Neto, Galvao and Natividade \cite{BarcelosNeto:1991dp} using the Batalin-Fradkin-Vilkovisky (BFV) approach
permitting a covariant gauge fixing.
Once again the original phase space considered contained conjugated momenta to the field velocities and accelerations.

With a fair comprehension of the gauge sector, from the beginning of this century on, the focus moved to the matter sector, concerning for instance renormalizability
\cite{Bufalo:2012tt}, scattering amplitude calculations \cite{Bufalo:2014jra}, radiative corrections \cite{Bufalo:2010sb} and finite temperature field theory \cite{Bonin:2009je}. 
In \cite{Bufalo:2012tt}, Bufalo, Pimentel and Zambrano considered the electromagnetic coupling of Podolsky's vector field with fermions, resulting in the so called
Generalized Quantum Electrodynamics (GQED${}_4$).  The standard renormalization program was followed with the respective counterterms calculations.  Additionally, from
the known experimental value of the electron magnetic moment, a numerical bound for Podolsky's parameter was obtained in \cite{Bufalo:2012tt}.
In \cite{Bufalo:2014jra}, Bufalo, Pimentel and Souto have calculated the generalized electrodynamics contribution to the Bhabha scattering cross section attempting
to address known discrepancies between usual QED predictions and experiment.

Regarding generalizations of classical Maxwell theory, particularly on the self-force of a charged
point particle in an electromagnetic field, Podolsky's model has been also revisited last year by Gratus, Perlick and Tucker \cite{Gratus:2015bea}.
These authors investigated in detail whether Podolsky's proposal can fix the known divergences related to the description of the general
motion of a charged particle in an electromagnetic field.

From the classical point of view, not long ago Kruglov formulated a first-order system of differential equations version for Podolsky's equations of motion \cite{Kruglov:2009yr}.
As a means of shedding more light in some of the current related issues above mentioned we suggest to consider deeper the physical interpretations of a reduced-order approach.
We discuss the introduction of only one extra vector field with a simple physical interpretation, obtaining a second-order system of differential equations with the
corresponding physical action and Hamiltonian.  We further proceed with the quantization process of the system through the canonical and functional integral analysis, obtaining
the generating functional at quantum level.

In this present article, since we are mostly interested in the gauge sector of Podolsky's generalized electrodynamics where the higher-order derivatives occur, we consider for simplicity only the electromagnetic field.
The inclusion of matter fields, either bosonic or fermionic, should pose no major problems.  Our work is organized as follows:
In section {\bf 2} we present Podolsky's original Lagrangian aiming to fix notation and conventions.
In section {\bf 3} we show that Podolsky's model is equivalent to a reduced-order derivatives one.  The price to be paid is the introduction of an additional
massive vector field.  We discuss and compare the corresponding differential equations for both models.
In section {\bf 4} we provide a Hamiltonian for the reduced-order model and determine its canonical structure of two first-class and two second-class constraints.
The quantization is then performed via Dirac brackets after gauge-fixing.  In section {\bf 5} we pursue the Senjanovic-BFV path integral quantization
introducing ghost fields for the gauge symmetry and obtaining the generating functional of the theory.
The last section is devoted to some concluding remarks.

\section{Podolsky's Generalized Electrodynamics}
Podolsky's original Lagrangian density, depending on a gauge vector field $A_\mu$, can be written as
\cite{Podolsky:1942zz}
\begin{equation}
\label{L0}
{\cal L}_P(A_\mu)
=
 -\frac{1}{4}F_{\mu\nu}F^{\mu\nu}
 +\frac{
 a^2
 }{2}\partial_\nu F^{\mu\nu}\partial^\rho F_{\mu\rho}
\,,
\end{equation}
where $F_{\mu\nu}$ is short for
\begin{equation}
F_{\mu\nu} = \partial_\mu A_\nu - \partial_\nu A_\mu
\,,
\end{equation}
and $a$ is an open length parameter.
The greek indices run from 0 to 3 and we use the Minkowski metric $(+,-,-,-)$.
This is a very simple Lagrangian indeed, the only roundabout being its second-order derivative 
term proportional to the parameter $a$.  Actually, since Podolsky's term depends on $A_\mu$ only through the combination $F_{\mu\nu}$, the model enjoys the same
gauge invariance as ordinary electrodynamics.
That is to say (\ref{L0}) remains invariant under the transformation
\begin{equation}\label{gt}
\delta A_\mu = \partial_\mu \Lambda
\end{equation}
for an arbitrary given function $\Lambda(x)$.

By demanding stationarity of the action
\begin{equation}\label{SP}
S_P=\int d^4x\, {\cal L}_P
\end{equation}
with respect to arbitrary variations of the gauge field $A_\mu$ we obtain the fourth-order equations of motion
\begin{equation}\label{EL}
(1+a^2\square)\partial_\mu F^{\mu\nu}=0\,,
\end{equation}
or equivalently
\begin{equation}\label{4oA}
  (1+a^2\Box)
  \left(
  \Box\eta^{\mu\nu}-\partial^\mu\partial^\nu
  \right)
  A_\nu=0\,.
\end{equation}
As usual the symbol $\square$ denotes the D'Alembertian operator here given by $\square = \partial_\mu\partial^\mu$.
Note that the equations of motion are also invariant under (\ref{gt}). 
We can understand (\ref{EL}) as Podolsky's version of Maxwell equations -- for $\nu = 0$ we have Gauss' law and for
spatial values of the index $\nu$ we get the Maxwell-Amp\`ere equation.

It is clear that the model (\ref{L0}) describes a neat generalized electrodynamics reproducing ordinary Maxwell's theory in
the limit $a\rightarrow 0$.  As a result, which can be seen already from the equations of motion, the gauge field acquires a new $1/a$ mass
in a gauge invariant way, besides the usual massless mode.
However, we point out that the limit $a\rightarrow 0$ is not a continuous one with respect to the degrees of freedom of the model as
it involves an abrupt change in the constraint structure of the theory.
Curiously enough the number of degrees of freedom of Podolsky's model for $a\neq0$ is five.  We shall clarify this point further after elaborating
on the Hamiltonian analysis and constraints classification in the next section.

If we fix a gauge and directly quantize model (\ref{L0}), we are forced to deal with a vector particle with two possible mass excitations
corresponding to zero and $1/a$ poles of the propagator.  An alternative interpretation to be shown in the next section is to consider
these two distinct massive models as associated to two different fields.

\section{Reduced Order Model}
In order to 
study the time evolution
of the model (\ref{L0}) 
and proceed to quantization
we need to write down the corresponding Hamiltonian in
phase space.  It happens that, due to gauge invariance, the Lagrangian (\ref{L0}) is singular leading to a constrained Hamiltonian
system.  A straightforward constraint structure analysis has been performed in \cite{Galvao:1986yq} and reviewed in \cite{BarcelosNeto:1991dp} where it has been necessary to introduce canonical
momenta conjugated both to $A_\mu$ as well as to its time derivative ${\dot A}_\mu$.  Aiming to avoid the necessity of considering $A_\mu$ and ${\dot A}_\mu$
as independent fields here we propose the alternative path of first reducing the derivatives order of (\ref{L0}) by means of introducing
an auxiliary vector field $B_\mu$.
Instead of (\ref{L0}) we write down the Lagrangian density
\begin{equation}\label{L}
{\cal L}[A_\mu, B_\mu]
= -\frac{1}{4}F_{\mu\nu} F^{\mu\nu}
-\frac{a^2}{2}B_\mu B^\mu
+a^2\partial_\mu B_\nu F^{\mu\nu}
\end{equation}
and consider the
corresponding 
reduced-order action $S=\int d^4 x {\cal L}$. 
Similarly to (\ref{SP}) the reduced action is also gauge invariant under
\begin{equation}
\delta A_\mu = \partial_\mu \Lambda\,,\,\,\,\,\,\delta B_\mu = 0\,.
\end{equation}
Now the variation of this reduced-order action with respect to the fields $A_\mu$ and $B_\mu$ leads to the coupled equations of motion
\begin{equation}\label{Aeq}
0=
\frac{\delta S}{\delta A_\mu} 
=
(\square\eta^{\mu\nu} - \partial^{\mu}\partial^\nu)
(A_\nu - a^2 B_\nu)
\,,
\end{equation}
and
\begin{equation}\label{Beq}
0=
\frac{\delta S}{\delta B_\mu} 
=
a^2(\partial_\nu F^{\mu\nu} - B^\mu)
\,.
\end{equation}
This represents a system of eight linear partial differential equations of second-order on the fields $A_\mu$ and $B_\mu$ which is
equivalent to the four fourth-order ones (\ref{4oA}) on the $A_\mu$ field.  This can be explicitly seen when we substitute the relation
$B^\mu = \partial_\nu F^{\mu\nu}$
from the second set above
into the first set (\ref{Aeq}) reproducing the initial system (\ref{4oA}).

Another easy way to see it is the following:  Starting from equations (\ref{4oA}) define the auxiliary field
$B^\mu = -(\square\eta^{\mu\nu}-\partial^\mu\partial^\nu)A_\nu$.  Then on the one hand the fourth-order set (\ref{4oA}) is clearly
equivalent to the second-order set
\begin{equation}\label{Bwave}
(1+a^2\square)B^\mu = 0 
\,,
\end{equation}
and
\begin{equation}\label{Awave}
(\square\eta^{\mu\nu}-\partial^\mu\partial^\nu)A_\nu = - B^\mu
\,.
\end{equation}
On the other hand the equations (\ref{Awave}) are nothing more than (\ref{Beq}) while subtracting (\ref{Bwave}) from (\ref{Awave}) and
considering that $\partial_\mu B^\mu = 0$ reproduces (\ref{Aeq}).

Therefore the two systems of partial differential equations are equivalent.  Furthermore, from the physical point of view, we observe
that the auxiliary field $B_\mu$ is divergenceless and satisfies a Klein-Gordon equation (\ref{Bwave}) with mass $1/a$.
We can safely say that the $A_\mu$ field remains massless while the $1/a$ mass vector excitations have been transferred to the $B_\mu$ field.
It is interesting to remark here that a duality relation between the Podolsky and Proca models involving modifications in the mass term sign
has been recently pointed out in \cite{Abreu:2010zza}.

We claim that the Lagrange densities (\ref{L0}) and (\ref{L}) describe the same physical system at classical and quantum levels.  In the
next two sections we analyze the quantization of (\ref{L}) both using canonical as well as functional path integral methods
and compare our results with those of (\ref{L0}) established and published in the recent literature.

\section{Constraint Structure and Canonical Quantization}
Both models (\ref{L0}) and (\ref{L})  describe singular systems from the Dirac-Bergmann formalism \cite{Dirac, Dirac:1950pj, Bergmann} point of view in the sense
that their Hessian are null.  Additionaly, since they enjoy local gauge invariance we expect them to exhibit first-class
constraints in the phase space.  
The constraint structure of the original system (\ref{L0}) was analyzed
thoroughly in \cite{Galvao:1986yq} where it was shown that there are three constraints in phase space, all of them first-class.
In our case, since we have isolated the gauge symmetry in the $A_\mu$ sector and have introduced the gauge invariant auxiliary
vector field $B_\mu$ we expect also second-class constraints.
In the following we apply the Dirac-Bergmann algorithm to system (\ref{L}) by going to the phase space and considering its instant-form
time evolution as a constrained system.  The quantization will then be achieved by fixing the gauge, calculating Dirac brackets, and sending
them to quantum operator commutators.

As a starting point,
in order to calculate the conjugated momenta and perform a Legendre transformation, we split the Lagrangian
(\ref{L}) into two parts with respect to the occurrance of time derivatives writing
\begin{equation}\label{Lt}
{\cal L}=
\frac{1}{2}F_{0i}F_{0i} - a^2(\partial_0 B_i - \partial_i B_0) F_{0i} - {\cal H}_{sp}
\,,
\end{equation}
with
\begin{equation}
{\cal H}_{sp}
\equiv
\frac{1}{4}F_{ij}F_{ij}-a^2\partial_i B_j F_{ij} + \frac{a^2}{2}B_\mu B^\mu
\,.
\end{equation}
This splitting is done just for operational convenience since the Legendre transformation deals only with terms containing time derivatives.  The
definition of ${\cal H}_{sp}$ will also turn out to be handy for notational purposes.
Associated to the fields $A_\mu$ and $B_\mu$ we introduce respectively the canonical momenta
\begin{equation}
\Pi^\mu = \frac{\partial {\cal L}}{\partial {\dot A}_\mu}
\,,
\end{equation}
and
\begin{equation}
\Pi_B^\mu = \frac{\partial {\cal L}}{\partial {\dot B}_\mu }
\,.
\end{equation}
As an immediate consequence of the very definition of the canonical momenta we have two primary constraints, namely,
\begin{equation}
\chi_1 =
\Pi_B^0
\,,
\end{equation}
and
\begin{equation}
\chi_3 =
\Pi^0
\,.
\end{equation}
Now the canonical Hamiltonian can be calculated as the usual Legendre transformation of (\ref{Lt}), leading to
\begin{equation}\label{Hc}
H_c=\int d^3x \left[
-\frac{\Pi^i\Pi_B^i}{a^2}
-\frac{\Pi^i_B\Pi^i_B}{2a^4}
+{\cal H}_{sp}
-A_0\partial_i\Pi^i
-B_0\partial_i\Pi^i_B
\right]
\,.
\end{equation}
As usual, the canonical Hamiltonian is well defined only within the primary constraint surface.
It is interesting to compare (\ref{Hc}) with the Hamiltonian obtained in \cite{Galvao:1986yq}.
Clearly there exists a natural correspondence beteween the momentum field $\pi^i$ conjutated to the accelerations in \cite{Galvao:1986yq} and our
$\Pi^i_B$, although the Hamiltonian in \cite{Galvao:1986yq} contains higher-order derivatives as compared to (\ref{Hc}).
Note also that here both $A_0$ and $B_0$ play roles of Lagrange multipliers enforcing secondary constraints, however, as previously anticipated,
$A_0$ refers to a massless field and $B_0$ to a $1/a$ mass field.
Once characterized the canonical Hamiltonian and primary constraints we may introduce two Lagrange multipliers $\lambda$ and $\lambda_B$ and
define the primary Hamiltonian as
\begin{equation}
H_P = H_c + \int d^3x
\left[
\lambda \Pi^0
+\lambda_B \Pi^0_B
\right]
\,.
\end{equation}
Proceeding with the Dirac-Bergmann algorithm, the
imposition of time conservation for the primary constraints generates two secondary constraints given by
\begin{equation}
\chi_2 = \partial_i\Pi^i_B-a^2B_0\,,
\end{equation}
and
\begin{equation}
\chi_4=\partial_i\Pi^i
\,.
\end{equation}
Further time conservation does not lead to new constraints, rather determines one of the Lagrange multiplier as
\begin{equation}
\lambda_B=\partial_i B_i
\,.
\end{equation}
The other Lagrange multiplier $\lambda$, associated to $\chi_3$, remains undetermined signalizing the presence of first-class constraints generating gauge
symmetries.  In fact we have obtained the whole set of constraints of the theory and, as can be easily checked, it turns out that $\chi_1$ and
$\chi_2$ are second-class while $\chi_3$ and $\chi_4$ are first-class.

Our analysis has shown a different constraint structure when compared to
\cite{Galvao:1986yq} where the instant-form evolution of (\ref{L0}) in phase space occurs under the regency of three first-class constraints.
No second-class constraints appear in \cite{Galvao:1986yq}.
However, the number of degrees of freedom of the two equivalent models is exactly the same, as it
should be.  In the case considered in \cite{Galvao:1986yq} there were a total of sixteen fields in phase space, namely
$(A_\mu,{\dot A}_\mu)$ and their corresponding momenta.  Subtracting three first-class constraints and three gauge conditions
we are left with ten independent fields in phase space implying a total of five degrees of freedom after dividing by two.
In the present case we have also sixteen fields in phase space, that is $(A_\mu, B_\mu)$ and corresponding momenta.  However,
from that amount we subtract two second-class constraints, two first-class constraints and two gauge fixing conditions.
After dividing by two again we are left with five degrees of freedom.

The interpretation of these five degrees of freedom becomes clearer in the current reduced-order equivalent model, two of them belong to the
massless gauge vector field $A_\mu$ and three to the massive invariant vector field $B_\mu$.
In Podolsky's original model, the sole gauge field $A_\mu$ contains both massless and $1/a$ massive excitations, here these two propagating modes
appear decoupled into two vector fields.

\begin{table}\label{DB}
\caption{Dirac Brackets}
\centering
\begin{tabular}{c|ccccc}
&$A_j$&$B_0$&$B_j$&$\Pi^j$&$\Pi^j_B$\\
\hline
$A_i$&.&.&.&$(\delta_i^j-\frac{\partial_i\partial_j}{\nabla^2})$&.\\
$B_0$&.&.&$-\frac{1}{a^2}\partial_j$&.&.\\
$B_i$&.&$-\frac{1}{a^2}\partial_i$&.&.&$\delta^j_i$\\
$\Pi^i$&$(-\delta^i_j+\frac{\partial_i\partial_j}{\nabla^2})$&.&.&.&.\\
$\Pi^i_B$&.&.&$-\delta^i_j$&.&.
\end{tabular}
\end{table}

Once we have the whole set of constraints at our disposal, the next step is to calculate the corresponding Dirac Brackets (DB) to be sent to the quantum commutators
upon quantization.  Aiming to obtain an invertible constraint matrix, we introduce two gauge fixing conditions for the first-class constraints.  To achieve the
radiation gauge we choose
\begin{equation}\label{gf}
\chi_5 = A_0\,,\,\,\,\,\,\,\chi_6 = \partial_i A_i\,.
\end{equation}
Note that in our approach we need fewer gauge conditions than in \cite{Galvao:1986yq}.
Actually, due to the different constraint structure of (\ref{L0}), in \cite{Galvao:1986yq} a total of three gauge fixing conditions was necessary
in the traditional higher derivatives approach.

Now we have all the ingredients to calculate the DB in phase space. The invertible constraint matrix $C_{\alpha\beta}$ 
considering the two gauge fixing conditions (\ref{gf}) for $\alpha, \beta = 1,\dots,6$,
reads 
\begin{equation}\label{C}
C_{\alpha\beta}
({\bf x},{\bf y})
 = 
\left(
\begin{array}{cccccc}
.&a^2&.&.&.&.\\
-a^2&.&.&.&.&.\\
.&.&.&.&-1&.\\
.&.&.&.&.&\partial_i\partial_i\\
.&.&1&.&.&.\\
.&.&.&-\partial_i\partial_i&.&.
\end{array}
\right)
\delta({\bf x}-{\bf y})
\end{equation}
with dots standing for null entries.
As a remark, for our notational convention, when not specified the derivatives of delta functions refer to the first argument.
For two given phase space functions $A$ and $B$ the DB is defined as
\begin{equation}
[A({\bf x}),B({\bf y})]^*=[A({\bf x}),B({\bf y})]-
\int d{\bf u} d{\bf v}\,
[A({\bf x}),\chi_\alpha({\bf u})]C^{\alpha\beta}({\bf u},{\bf v})[\chi_\beta({\bf v}),B({\bf y})]
\,,
\end{equation}
where $C^{\alpha\beta}$ denotes the inverse of (\ref{C}).
By convention, all brackets are calculated at equal times.
Table 1 shows the explicit results for the DB among 
some of
the fundamental phase space variables -- the dots stand for zero and a spatial Dirac delta
$\delta ({\bf x}-{\bf y})$ should be understood in all entries.
The phase space variables which do not show up in Table 1 have identically vanishing DB, namely
\begin{equation}
[A_0,X]^* = [\Pi^0,X]^* = [\Pi^0_B,X]^* = 0\,,
\end{equation}
where $X$ denotes any arbitrary phase space function.
Two specific important DB's from Table 1 we would like to mention are
\begin{equation}
[A_i({\bf x}),\Pi^j({\bf y})]^*= (\delta_i^j-\frac{\partial_i\partial_j}{\nabla^2})
\delta({\bf x}-{\bf y})
\end{equation}
and
\begin{equation}
[B_i({\bf x}),\Pi^j_B({\bf y})]^*=\delta_i^j\delta({\bf x}-{\bf y})
\end{equation} 
which confirm that only the gauge field $A_\mu$ is transverse in the radiation gauge while $B_\mu$ describes a species of massive Proca field.
The relation
\begin{equation}
[B_i({\bf x}),B_0({\bf y})]^*=-\frac{1}{a^2}\partial_i\delta({\bf x}-{\bf y})
\end{equation}
ensures that the DB of $\chi_2$ with any other phase space function identically vanishes.

A comparison of the DB structure obtained here with that presented in \cite{Galvao:1986yq} shows that we have been
able to achieve a more transparent and easier to understand result, due to the physical interpretation of the fields $A_\mu$ and $B_\mu$ in the
reduced-order model.
In particular the inherent nonlocality of the current 
DB structure
is of lower level than those of \cite{Galvao:1986yq} which rely on the inverse of a fourth-order
differential operator.

We have successfully calculated the Dirac brackets among the phase space variables for the model (\ref{L}).
From this point on the canonical quantization follows the usual procedure promoting the classical variables of phase space
to quantum operators and associating the DB's to the commutation relations.

\section{Functional Senjanovic-BFV Quantization}
In the last section we have calculated the DB structure for the reduced-order Podolsky model.
As it is well known however,
one of the disadvantages of the canonical quantization for gauge systems is the somewhat cumbersome nonlocal character of the Dirac Brackets.  
In the present case this can be seen in Table
1 where the inverse of the operator $\partial_i\partial_i$ plays a crucial role.
This is a natural consequence of the fact that in its original form the Dirac Bergmann algorithm, based on a Hamiltonian approach, is not
well suited for covariant gauges.
As an alternative to the canonical Dirac Bracket quantization, in this section, we discuss the path integral quantization in the Lorenz covariant gauge.  Since the system
possesses a mixed constraint structure we shall use the Senjanovic \cite{Senjanovic:1976br} approach for the second-class sector
and the Batalin Fradkin Vilkovisky (BFV) \cite{Fradkin:1975cq, Batalin:1977pb} for the gauge invariant first-class sector.
As mentioned in the introduction, the direct quantization of (\ref{L0}) by functional integral techniques was performed in \cite{BarcelosNeto:1991dp} where only first-class constraints were
considered. 

As a necessary first step for the BFV functional quantization we consider the extended phase space for the model (\ref{L}) constructed from
the original field variables and momenta
\begin{equation}
(A_\mu,\Pi^\mu)\,,\,\,\,\,\,\,(B_\mu,\Pi_B^\mu)\,,
\end{equation}
and the extra fields
\begin{equation}
(\lambda,b)\,,\,\,\,\,\,\,(C,{\bar{\cal P}})\,,\,\,\,\,\,\,({\bar C},{\cal P})\,.
\end{equation}
The latter stand for
the undetermined Lagrange multiplier $\lambda(x)$ associated to the primary first-class constraint $\chi_3$ and its conjugated momentum
field $b(x)$, the Grassmann field variables $C(x)$ and ${\bar C}(x)$ for symmetries generated by the first-class constraints and their corresponding anticommuting momenta
${\bar {\cal P}}(x)$ and ${\cal P}(x)$.   
The fundamental transition amplitude for the model (\ref{L}) can then be written as
\begin{equation}\label{Z}
Z=\int [d\nu]\,\exp (iS_{eff})
\end{equation}
with functional integration measure
\begin{equation}\label{measure}
[d\nu] = D
\!
A_\mu 
\,
D\Pi^\mu\, D
\!
B_\mu 
\,
D\Pi^\mu_B\, D
\!
\lambda 
\,
Db\, DCD{\bar C} D{\cal P}D{\bar{\cal P}}
\,||\det \{\chi_1,\chi_2\}||\,
\delta(\chi_1)\delta(\chi_2)
\end{equation}
and effective action
\begin{equation}
S_{eff} = \int d^4 x
\left\{
{\dot A}_\mu
\Pi^\mu + 
{\dot B}_\mu
\Pi^\mu_B  
+ {\dot \lambda} b
+ {\dot C}{\bar {\cal P}} 
+ {\dot{\bar C}}{\cal P}
- {\cal H}_{min}
+ [\Omega,\Psi]
\right\}
\,.
\end{equation}
For simplicity, in (\ref{Z}) and in its further developments, we do not include explicitly the external sources.
In the last equation
${\cal H}_{min}$, $\Omega$ and $\Psi$ represent respectively the minimal BRST invariant Hamiltonian density, the BRST charge and the gauge-fixing
fermion according to the usual BFV quantization scheme prescription \cite{Fradkin:1975cq, Batalin:1977pb}.  Considering that in our present case
the gauge algebra satisfies
\begin{equation}
[\chi_3,H_c] = \chi_4 
\,,
\end{equation}
we have
\begin{equation}
{\cal H}_{min}
=
{\cal H}_c + {\bar{\cal P}}{\cal P}
\,,
\end{equation}
with ${\cal H}_c$ being the canonical Hamiltonian density corresponding to (\ref{Hc}) and
\begin{equation}
\Omega
=\int d^3y
\left\{
{\cal P}\Pi^0+C\partial_i\Pi^i
\right\}
\,.
\end{equation}

The BRST charge $\Omega$ satisfies the relation
\begin{equation}
[\Omega,\Omega]=0
\end{equation}
and generates the BRST transformations in the extended phase space
\begin{equation}\label{BRST}
\delta A_0 = - \epsilon {\cal P}\,,\,\,\,\,\,\,
\delta A_i = \epsilon\partial_i C\,,\,\,\,\,\,\,
\delta{\bar C}= -\epsilon\Pi^0\,,\,\,\,\,\,\,
\delta{\bar P}= -\epsilon\partial_i\Pi^i\,,
\end{equation}
which leave $H_{min}\equiv\int d^3x\,{\cal H}_{min}$ invariant.
In equations (\ref{BRST}) $\epsilon$ denotes a Grassmann anticommuting parameter.
To achieve the standard Lorenz covariant gauge we choose the gauge fixing fermion as
\begin{equation}
\Psi = {\bar C}(\partial_i A_i - \frac{\xi}{2}\Pi^0)
+{\bar{\cal P}}\lambda
\end{equation}
where $\xi$ is an arbitrary real gauge-fixing parameter.
The generating functional (\ref{Z}) is now well defined in the extended phase space.  In the following, in order to achieve a simpler
explicitly covariant expression for (\ref{Z}) in configuration space, we shall perform the momenta field variables integration.  First, to ensure the second
class constraints, we integrate in $\Pi^0_B$ using the Dirac delta functional $\delta({\chi}_1)$ and introduce an auxiliary field variable
$\Gamma(x)$ to write
\begin{equation}
\delta(\chi_2)=\int D\Gamma\, e^{i\Gamma(\partial_i\Pi^i_B-a^2B_0)}
\end{equation}
in the measure (\ref{measure}).
After a change of variables $B_0\rightarrow B_0-\Gamma$ the integrations in $\Pi^i_B$ and $\Gamma$ can also be done leading to
the partial result
\begin{equation}
Z =
\int [d\nu']\,\exp (i{S'})
\end{equation}
with a shorter integration measure $[d\nu']$ up to a non relevant constant factor given by
\begin{equation}
[d\nu']=
D
\!
A_\mu 
\,
D\Pi^\mu\, D
\!
B_\mu 
D
\!
\lambda 
\,
Db\, DCD{\bar C} D{\cal P}D{\bar{\cal P}}
\end{equation}
and effective partial action
\begin{equation}
\begin{array}{lcl}
S'&=&
{\displaystyle
\int} d^4 x
\left\{
{\dot A}_\mu 
\Pi^\mu
+  {\dot \lambda} b
+ {\dot C}{\bar {\cal P}}
+ {\dot{\bar C}}{\cal P}
-\frac{a^4}{2}
{\left(
\partial_0 B_i - \partial_i B_0 + \frac{\Pi^i}{a^2}
\right)}^2
\right.\\&&\left.
-{\cal H}_{sp}
+ A_0\partial_i\Pi^i
-{\bar{\cal P}}{\cal P}
+ [\Omega,\Psi]
                                                   \vphantom{\left(\partial_0 B_i - \partial_i B_0 + \frac{\Pi^i}{a^2}\right)}
\right\}
\,.
\end{array}
\end{equation}
We have also reabsorbed the term $||\det \{\chi_1,\chi_2\}||$, which is proportional to $a^2$, in the integration measure.

To continue further integrating in the momenta fields we perform the change of variables $A_0\rightarrow A_0 +\lambda$.
That turns the integrations in $b$ and $\lambda$ easy leading to a constant factor which can also be absorbed in the integration measure.
After these steps we may rewrite the generating functional as
\begin{equation}
Z = \int [d\nu'']\, \exp(iS'')
\end{equation}
with integration measure
\begin{equation}
[d\nu''] =
D
\!
A_\mu 
\,
D\Pi^\mu\, D
\!
B_\mu 
\, 
DCD{\bar C} D{\cal P}D{\bar{\cal P}}
\end{equation}
and a more pleasant, almost final, action
\begin{equation}
\begin{array}{lcl}
S''&=&
{\displaystyle
\int} d^4 x
\left\{
{\dot A}_\mu 
\Pi^\mu
+ {\dot C}{\bar {\cal P}}
+ {\dot{\bar C}}{\cal P}
-\frac{a^4}{2}
{\left(
\partial_0 B_i - \partial_i B_0 + \frac{\Pi^i}{a^2}
\right)}^2
-{\cal H}_{sp}
\right.\\&&\left.
+ A_0\partial_i\Pi^i
-{\bar{\cal P}}{\cal P}
-\Pi^0\partial_i A_i +\frac{\xi}{2}\Pi^0\Pi^0
-{\bar C}\partial_i\partial_i C
                                                   \vphantom{\left(\partial_0 B_i - \partial_i B_0 + \frac{\Pi^i}{a^2}\right)}
\right\}
\,.
\end{array}
\end{equation}

Finally we perform the integrations in $\Pi^\mu$, $\cal P$ and $\dot {\cal P}$.  Specifically the integration in $\Pi^0$ brings the gauge
fixing term containing ${(\partial_\mu A^\mu)}^2$ down to the argument of the exponential, the integration in $\Pi^i$ provides the term coupling the vector fields and the integration
in the ghost momenta the factor ${\dot C}{\dot{\bar C}}$ which is added to the similar spatial term to produce a corresponding
covariant explicitly term.  That said and done the final expression for the generating functional in terms of a covariant action in configuration
space reads
\begin{equation}\label{Zf}
Z=\int [d\mu] \exp(iS)
\end{equation}
with
\begin{equation}\label{df}
[d\mu] = D\!A_\mu\,D\!B_\mu\,DC D{\bar C}
\end{equation}
and
\begin{equation}\label{Sf}
S=\int d^4x
\left\{
-\frac{1}{4}F_{\mu\nu} F^{\mu\nu}
-\frac{a^2}{2}B_\mu B^\mu
+a^2\partial_\mu B_\nu F^{\mu\nu}
-\frac{1}{2\xi}{(\partial_\mu A^\mu)}^2
+{\bar C}\partial_\mu\partial^\mu C
\right\}
\,.
\end{equation}
This ends the Senjanovic-BFV quantum analysis of the model (\ref{L}), considering the first and second-class constraints. It is now clear that the the auxiliary vector
field $B_\mu$ is only related to the second-class sector being decoupled from the gauge sector.

\section{Conclusion}

With the introduction of an auxiliary massive vector field $B_\mu$ we reduced the order of the derivatives present in Podolsky's generalized electrodynamics and pursued
the quantization of the resulting model.  The dynamical evolution was shown to occur in a phase space containing two first-class and two second-class constraints,
with a total number of five degrees of freedom, the same as the usual higher-order model. The quantization was done in two distinct forms, namely
via the canonical approach in the radiation gauge and via functional quantization in the Lorenz gauge.  Regarding the canonical
quantization we calculated the Dirac bracket structure of the theory involving both vector fields $A_\mu$ and $B_\mu$.
For the path integral analysis we implemented the second-class constraints directly in the integration measure via the Senjanovic procedure.  The gauge sector 
had to be more carefully handled by means of extending the phase space with ghost fields and pursuing the BFV quantization.

The whole analysis has shown that Podolsky's model can be considered equivalent to the reduced order one (\ref{Zf}-\ref{Sf}) at classical and quantum levels.
The massive propagating modes of the original Podolsky's model can be viewed as associated to the auxiliary vector field.
It is clear from the final action (\ref{Sf}) that $B_\mu$ behaves as a Proca field with the opposite mass signal.  Due to its mass, the $B_\mu$
field does not transform under the BRST symmetry.

\textbf{Acknowledgments:} The author thanks Prof. Chueng-Ryong Ji for 
enlightening physical discussions on the subject and gratefully appreciates kind hospitality from the Physics Department at North Carolina State University during
postdoctoral research visit which made this work possible.  Financial support from the Brazilian funding agency {\it Conselho Nacional de Desenvolvimento Cient\'\i fico e Tecnol\'ogico - CNPq} under process number 202141/2015-2 is acknowledged.


\begin{thebibliography}{15}

\bibitem{Podolsky:1942zz}
  B.~Podolsky,
  Phys.\ Rev.\  {\bf 62}, 68 (1942).

\bibitem{Podolsky:1944zz}
  B.~Podolsky and C.~Kikuchi,
  Phys.\ Rev.\  {\bf 65}, 228 (1944).

\bibitem{Podolsky:1948}
  B.~Podolsky and P.~Schwed,
  Rev.\ Mod.\ Phys. {\bf 20}, 1 (1948).

\bibitem{Bufalo:2012tt} 
  R.~Bufalo, B.~M.~Pimentel and G.~E.~R.~Zambrano,
  Phys.\ Rev.\ D {\bf 86}, 125023 (2012).

\bibitem{Bufalo:2013joa} 
  R.~Bufalo and B.~M.~Pimentel,
  Phys.\ Rev.\ D {\bf 88}, no. 6, 065013 (2013).

\bibitem{Zayats:2013ioa} 
  A.~E.~Zayats,
  Annals Phys.\  {\bf 342}, 11 (2014).


\bibitem{Bufalo:2014jra} 
  R.~Bufalo, B.~M.~Pimentel and D.~E.~Soto,
  Phys.\ Rev.\ D {\bf 90}, no. 8, 085012 (2014).

\bibitem{Gratus:2015bea} 
  J.~Gratus, V.~Perlick and R.~W.~Tucker,
  J.\ Phys.\ A {\bf 48}, no. 43, 435401 (2015).

\bibitem{Barone:2016pyy} 
  F.~A.~Barone and A.~A.~Nogueira,
  Int.\ J.\ Mod.\ Phys.\ Conf.\ Ser.\  {\bf 41}, 1660134 (2016).



\bibitem{Cuzinatto:2011zz} 
  R.~R.~Cuzinatto, C.~A.~M.~de Melo, L.~G.~Medeiros and P.~J.~Pompeia,
  Int.\ J.\ Mod.\ Phys.\ A {\bf 26}, 3641 (2011).



\bibitem{Galvao:1986yq}
  C.~A.~P.~Galvao and B.~M.~Pimentel,
  Can.\ J.\ Phys.\  {\bf 66}, 460 (1988).


\bibitem{BarcelosNeto:1991dp}
  J.~Barcelos-Neto, C.~A.~P.~Galvao and C.~P.~Natividade,
  Z.\ Phys.\ C {\bf 52}, 559 (1991).


\bibitem{Bufalo:2010sb} 
  R.~Bufalo, B.~M.~Pimentel and G.~E.~R.~Zambrano,
  Phys.\ Rev.\ D {\bf 83}, 045007 (2011).

\bibitem{Bonin:2009je} 
  C.~A.~Bonin, R.~Bufalo, B.~M.~Pimentel and G.~E.~R.~Zambrano,
  Phys.\ Rev.\ D {\bf 81}, 025003 (2010).



\bibitem{Kruglov:2009yr} 
  S.~I.~Kruglov,
  J.\ Phys.\  {\bf 43}, 245403 (2010).




\bibitem{Abreu:2010zza} 
  E.~M.~C.~Abreu, A.~C.~R.~Mendes, C.~Neves, W.~Oliveira, C.~Wotzasek and L.~M.~V.~Xavier,
  Mod.\ Phys.\ Lett.\ A {\bf 25}, 1115 (2010).




\bibitem{Dirac}
 P. A. M. Dirac, {\it Lectures
in Quantum Mechanics}, Belfer Graduate School of Science, Yeshiva University
Press,  New York, (1964). 

\bibitem{Dirac:1950pj}
  P.~A.~M.~Dirac,
  Can.\ J.\ Math.\  {\bf 2} (1950) 129.

\bibitem{Bergmann} J. L. Anderson and P. G. Bergmann, Phys. Rev. \textbf{83 }%
(1951) 1018.

\bibitem{Senjanovic:1976br} 
  P.~Senjanovic,
  Annals Phys.\  {\bf 100}, 227 (1976).

\bibitem{Fradkin:1975cq} 
  E.~S.~Fradkin and G.~A.~Vilkovisky,
  Phys.\ Lett.\ B {\bf 55}, 224 (1975).

\bibitem{Batalin:1977pb} 
  I.~A.~Batalin and G.~A.~Vilkovisky,
  Phys.\ Lett.\ B {\bf 69}, 309 (1977).




\end{thebibliography}
\end{document}